\documentclass[a4paper,
              ]{jacow}

\makeatletter%                           % test for XeTeX where the sequence is by default eps-> pdf, jpg, png, pdf, ...
\ifboolexpr{bool{xetex}}                 % and the JACoW template provides JACpic2v3.eps and JACpic2v3.jpg which might generates errors
 {\renewcommand{\Gin@extensions}{.pdf,%
                    .png,.jpg,.bmp,.pict,.tif,.psd,.mac,.sga,.tga,.gif,%
                    .eps,.ps,%
                    }}{}
\makeatother

\ifboolexpr{bool{xetex} or bool{luatex}} % test for XeTeX/LuaTeX
 {}                                      % input encoding is utf8 by default
 {\usepackage[utf8]{inputenc}}           % switch to utf8

\usepackage[USenglish]{babel}

\ifboolexpr{bool{jacowbiblatex}}%        % if BibLaTeX is used
 {%
  \addbibresource{jacow-test.bib}
  \addbibresource{biblatex-examples.bib}
 }{}

\usepackage{epstopdf}

\begin{document}

\title{Study of Short Bunches at the Free Electron Laser CLIO\thanks{Work supported by the French ANR (contract ANR-12-JS05-0003-01), the IDEATE International Associated Laboratory (LIA) between France and Ukraine  and  Research Grant \#F58/380-2013 (project F58/04) from the State Fund for Fundamental Researches of Ukraine in the frame of the State key laboratory of high energy physics." }}

\author{ Nicolas Delerue\thanks{delerue@lal.in2p3.fr}, St\'ephane Jenzer,  Vitalii Khodnevych\textsuperscript{1}\thanks{hodnevuc@lal.in2p3.fr}
\\ LAL, Univ. Paris-Sud, CNRS/IN2P3, Universit\'e Paris-Saclay, Orsay, France.\\ 
Jean-Paul Berthet, Francois Glotin, Jean-Michel Ortega,  Rui Prazeres 
\\ CLIO/ELISE/LCP, Univ. Paris-Sud, CNRS, Universit\'e Paris-Saclay, Orsay, France.\\
 	 \textsuperscript{1}also at National Taras Shevchenko University of Kyiv, Kyiv, Ukraine\\
  }

\maketitle

\begin{abstract}
CLIO is a Free Electron Laser based on a thermionic electron gun. In its normal operating mode it delivers electron 8 pulses but studies are ongoing to shorten the pulses to about 1 ps. We report on simulations showing how the pulse can be shortened and the expected signal yield from several bunch length diagnostics (Coherent Transition Radiation, Coherent Smith Purcell Radiation).
\end{abstract}

\section{Introduction}
Experimental comparison of Coherent Smith-Purcell Radiation (CSPR)  and Coherent Transition Radiation (CTR) will take place at CLIO accelerator~\cite{clio}. This accelerator can produce single electron bunches  with length of about 8~$\pm1$~ps~\cite{blm}  and energy up to 50~MeV.  To predict the spectrums of CTR and CSPR and optimise the CLIO parameters for the experiment, we have performed simulations of  of the accelerator using ASTRA~\cite{astra}. 

\section{The CLIO accelerator}

\begin{figure*}[!bth]
    \centering
    \includegraphics*[width=\textwidth]{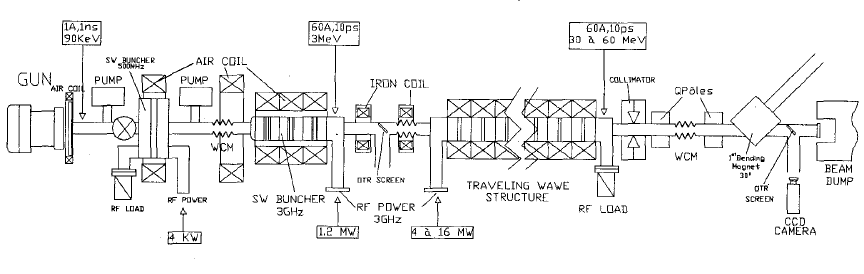}
    \caption{Layout of the CLIO accelerator (taken from \cite{optim}).}
    \label{clio}
%    \vspace*{-\baselineskip}
\end{figure*}

The CLIO free electron laser is an accelerator built in 1992. It is described in details in~\cite{clio} and it is shown on figure~\ref{clio}.
The CLIO accelerator consist of a thermionic gun, a subharmonic buncher (SHB), a fundamental buncher (FB) and an accelerating cavity (AC). The gun produce bunches about 1.5 ns long at an energy of 90 keV. These bunch are then compressed by the subharmonic buncher  to 200 ps or less  to make it suitable for further compression with the fundamental buncher. This fundamental buncher further compresses the beam to a few ps and accelerates bunch to several MeV, making the electrons relativistic. The bunches are then further accelerated in the accelerating cavity to the operation energy  (typically 15-50 MeV). \par

%\begin{figure}[!htb]
%  \centering
%  \includegraphics[width=0.9\linewidth]{plots/CLIO.png}
%  \caption{Layout of the CLIO accelerator (taken from \cite{optim}) XXX over two columns}
%  \label{clio}
%\end{figure}%

For our purpose we have simulated this accelerator using the ASTRA code. These simulations have been compared to the PARMELA~\cite{PARMELA} simulations available in the CLIO TDR~\cite{RT} and other documents~\cite{thesis-francois-glottin}.
The code ASTRA tracks particles through user defined external fields tacking into account the space charge field of the particle bunch. The tracking is based on a non-adaptive Runge-Kutta integration of 4th order. It is described in~\cite{astra}

%For ASTRA simulation longitudinal on-axis field $B_z$  is important.  The transverse field components is calculated   from the derivatives of the   on-axis field \cite{astra}. \par 

%Profile of magnetic field presented on the fig. \ref{fig:B1}.\par
%\begin{figure}[htb]
%  \centering
%  \includegraphics[width=0.9\linewidth]{plots/TransverseXY.eps}
 % \caption{Magnet field along the CLIO accelerator and change of the transverse bunch width}
%  \label{fig:B1}
%\end{figure}

The gun is a classical Pierce gridded gun with a thermoelectric dispenser cathode~\cite{comm}. This gun has a complicated geometry, so in the ASTRA simulation only a simplified model was used. All output parameters (emittance, transvserse distribution, bunch length, energy etc) from this simplified model were checked against the original data.\par 
The subharmonic buncher (SHB) is a stainless-steel reentrant cavity in the mode TM01 at 499.758 MHz i.e. the 1/6th subharmonic of the fundamental frequency of the accelerating cavity~\cite{optim}.

\begin{figure}[htb]
  \centering
  \includegraphics[width=0.9\linewidth]{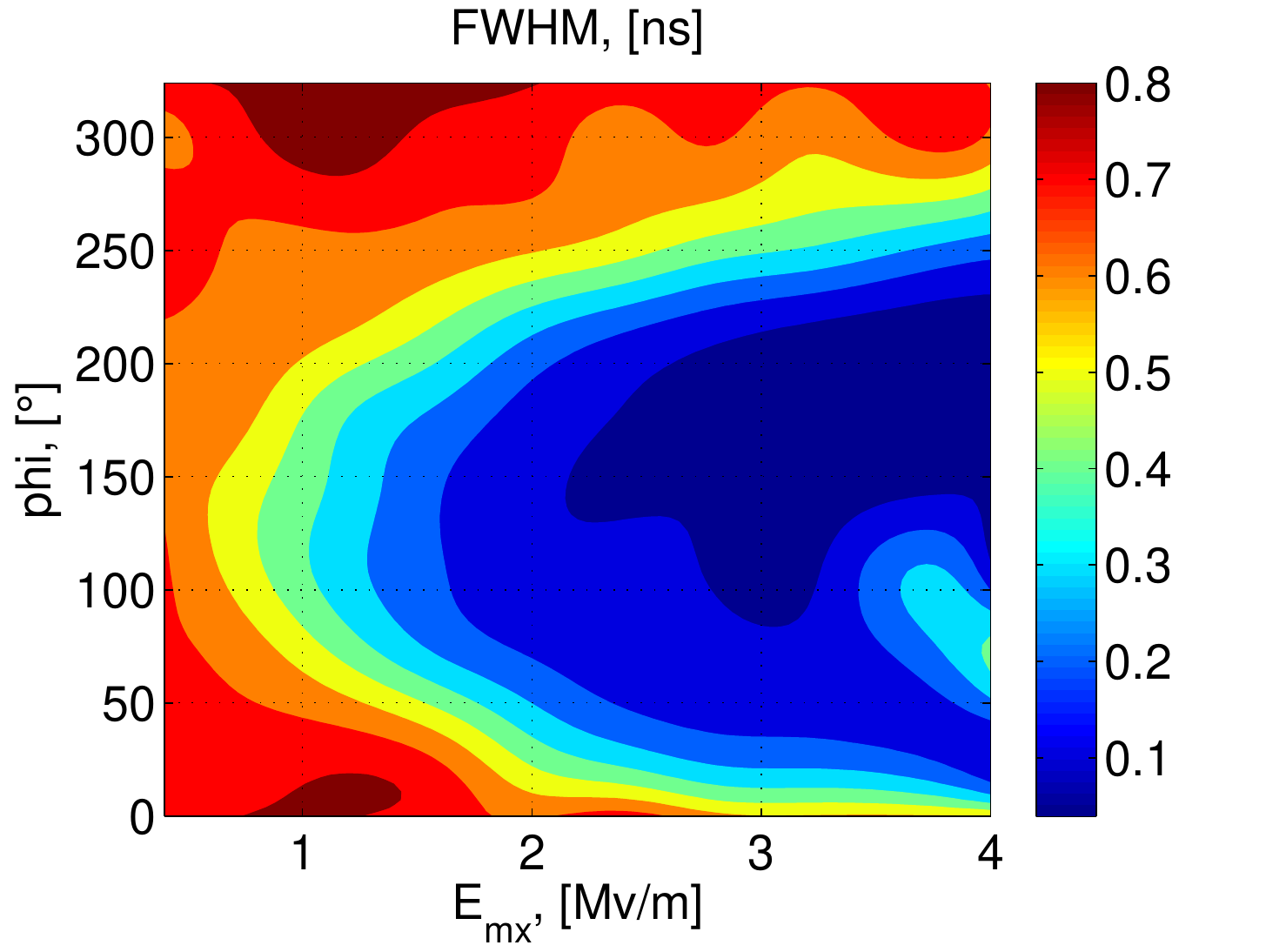}
  \caption{Full width at half  of maximum of the bunch for the sub-harmonic buncher phase and maximum field.}
    \label{shwdth}
\end{figure}

The phase of the accelerating cavity and the maximum field in that cavity play an important role in the final bunch length. To optimise these parameters we have used a 2D scan to find the optimal bunch properties. Figure~\ref{shwdth} shows the bunch Full-Width Half-Maximum (FWHM) as  a function of the phase and the field strength. \par

\begin{figure}[htb]
  \centering
  \includegraphics[width=0.9\linewidth]{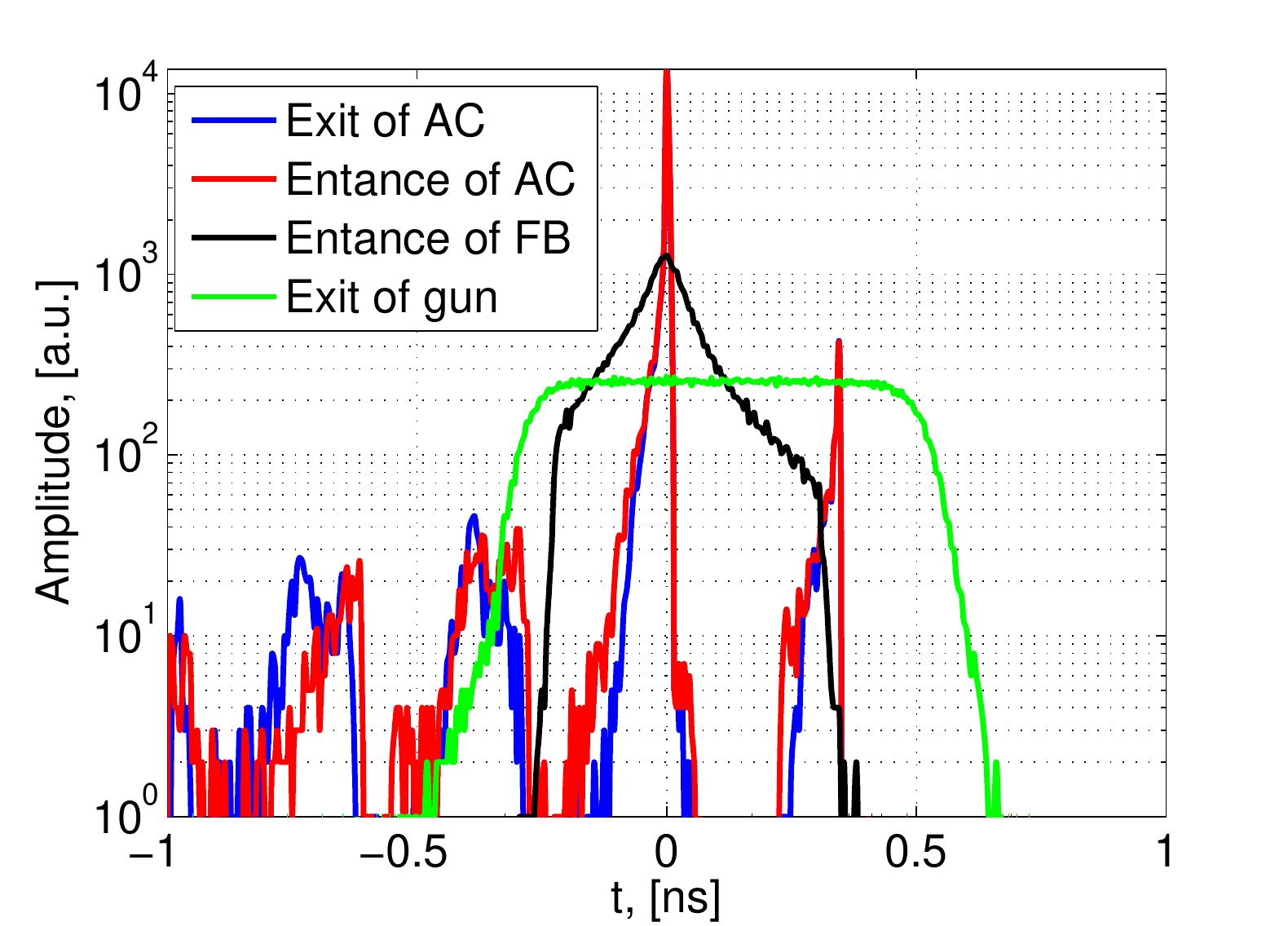}
  \caption{Longitudinal bunch size at exit of gun (FWHM=800 ps), at the entrance of FB (FWHM=92 ps), at the entrance (FWHM=2.35ps) and at the exit (FWHM=2.29ps) of the AC.}
    \label{dob1}
\end{figure}

Taking into account other parameters (charge in a 200ps window, full width at 10\% of the maximum, energy from the technical report \cite{RT}) we are able to choose the optimal values of the energy gradient and the phase of the cavity for the  SHB, 2.56 MV/m and 126 degrees respectively.  As  from the exit of the SHB to the entrance of the FB the bunch is still evolving, we look at the result of compression at the entrance of the FB. A comparison of the longitudinal bunch size at the gun's exit and at the entrance of the FB is presented on fig.~\ref{dob1}.\par 

The fundamental buncher (FB) is a copper triperiodic, S-band standing wave structure~\cite{LIL-cavity}. It is composed of cells at 3 different wavelengths, slightly matched to the beam velocity (0.92, 0.98 and 1 lambda)~\cite{clio}. The role of the 3 GHz buncher is to complete the compression of the pulses initiated by the SHB at 500~MHz and to bring the particles to ultra-relativistic energies~\cite{RT}.\par
 Similarly to the SHB, the FB also requires a phase study. We found the optimal phase to be 210 degree and the maximum cavity field as 22 MV/m. \par
The accelerating cavity (AC) is a constant gradient S band travelling wave disk-loaded structure. The cavity is surrounded with a set of solenoidal coils which give a continuous axial field adjustable up to 0.2 Tesla \cite{clio}. Comparison of the longitudinal bunch size at the entrance and at the exit of the AC is presented on the fig.~\ref{dob1}. In ideal conditions the profile is almost not altered in the AC, but particles are significantly accelerated.\par

%The result of the optimisation of the phase and field of all the components of the accelerator is shown on figure~\ref{p123}.

%energy distribution as on fig.~\ref{fig:energ} (
%

%\begin{figure}[!htb]
%  \centering
%  \includegraphics[width=0.9\linewidth]{plots/Profile20deg.eps}
%  \caption{}
%  \label{Prof1}
%\end{figure}%

%\begin{figure}[!htb]
%  \centering
%  \includegraphics[width=0.9\linewidth]{plots/Energy1.eps}
%  \caption{Energy distribution of the bunch}
%  \label{fig:energ}
%\end{figure}%

\begin{figure}[!htb]
  \centering
  \includegraphics[width=0.9\linewidth]{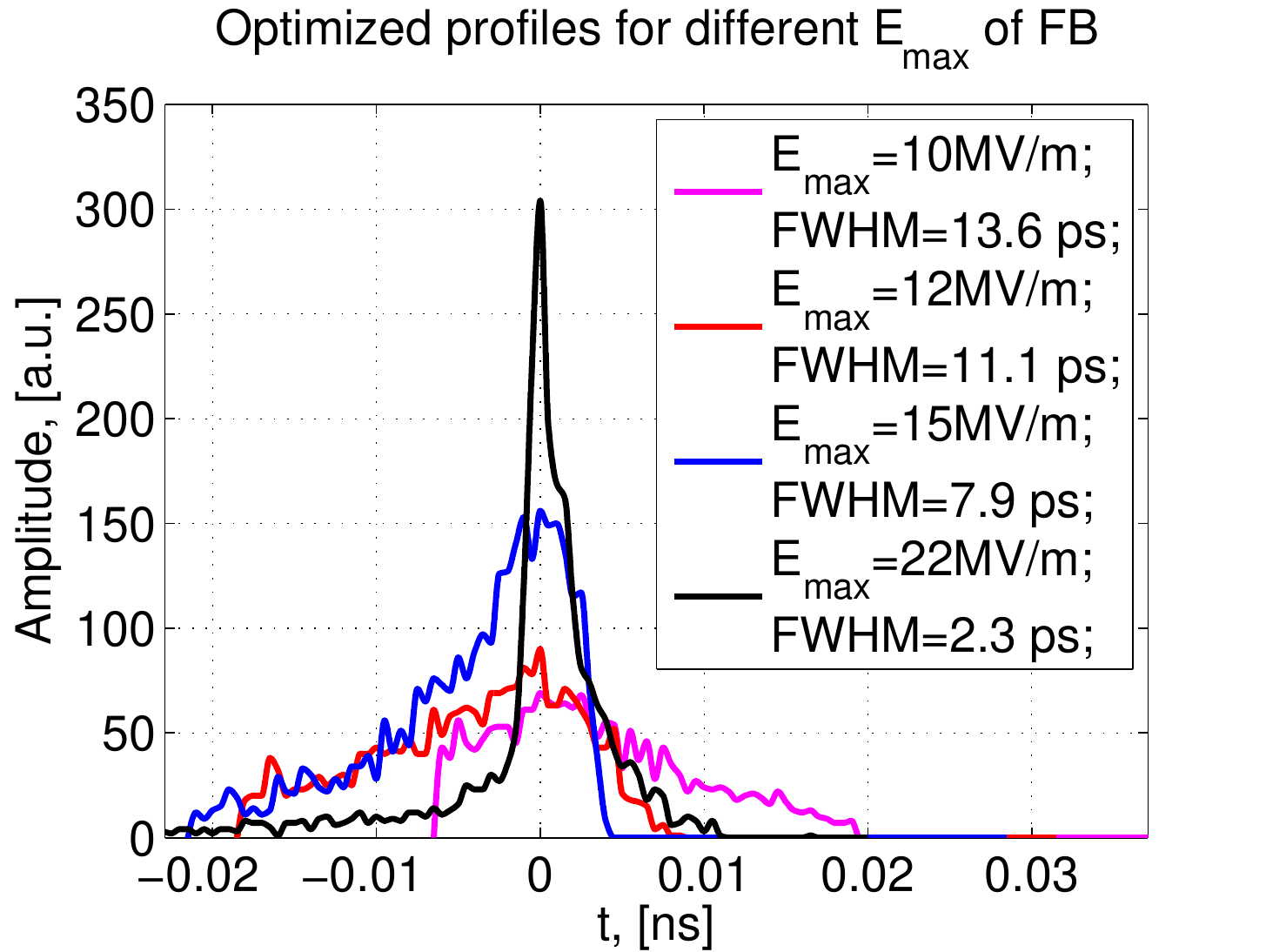}
  \caption{Profile of the bunch at the exit of the acceleration cavity for different maximum fields of fundamental buncher (optimized).}
  \label{p123}
\end{figure}

From our simulations we find that the parameter that has the most significant impact on the bunch length is the field in the FB (see fig.~\ref{p123}). The FB has been upgraded with respect to the original CLIO design and this explains why we can predict shorter bunches than what was originally foreseen in~\cite{blm,thesis-francois-glottin,comm}. Once this field is increased the phases of the other components need to be slightly optimized. To help us with this optimization we will use CTR and CSPR signals measured at the exit of the AC.

Our simulations also show that  it is possible to make even shorter bunches with the accelerator, but at the expense of degrading others parameters (energy spread, etc.). \par

%Some profiles in such conditions are shown on figure~\ref{p3}.\par
%\begin{figure}[!htb]
%  \centering
%  \includegraphics[width=0.9\linewidth]{plots/Profile.eps}
%  \caption{
%  Longitudinal profile of the bunch at the exit of the acceleration cavity after optimization. The profile presented here is far an energy of 60.34~MeV with an energy spread $\Delta\gamma/\gamma$=0.5\%.
%  Profile of the bunch at the exit of the acceleration cavity for different phases of the accelerating cavity.}
%  \label{p3}
%\end{figure}%

\section{Bunch length measurements}

To measure and optimize the bunch length at CLIO we plan to use install a bunch length monitor at the exit of the AC at CLIO. This bunch length monitor will use two different radiative phenomenon:  Coherent Smith-Purcell Radiation (CSPR)~\cite{CSPR} and Coherent Transition Radiation (CTR)~\cite{CTR}. A  comparison of CSPR  and CTR is presented in another contribution to this conference~\cite{MOPMB003}. \par

%These two phenomenon both rely on radiation that become more intense when the bunch length is sufficiently short with respect to the observed wavelength. It is therefore important to calculate the (incoherent) Single Electron Yield (SEY) for each of them and the form factor of the bunch. In both cases the radiation emitted will be of form:
%$$
%I_{\mbox{coh}}(\lambda) = I_1 ( N + N^2 \cal{F}(\lambda)  )
%$$
%where $I_{\mbox{coh}}(\lambda)$ is the total radiation emitted at wavelength $\lambda$, $I_1$ is the SEY, $N$ is the bunch charge (number of electrons) and $\cal{F}(\lambda)$ is the bunch form factor at $\lambda$. More details can be found in \cite{CSPR,CTR} and in references therein. The signal emitted in the bunch length monitor will be measured by pyroelectric detectors.
%A discussion of the SEY of CSPR  and CTR is presented in another contribution to this conference~\cite{MOPMB003}. \par

\begin{figure}[!htb]
  \centering
  \includegraphics[width=0.85\linewidth]{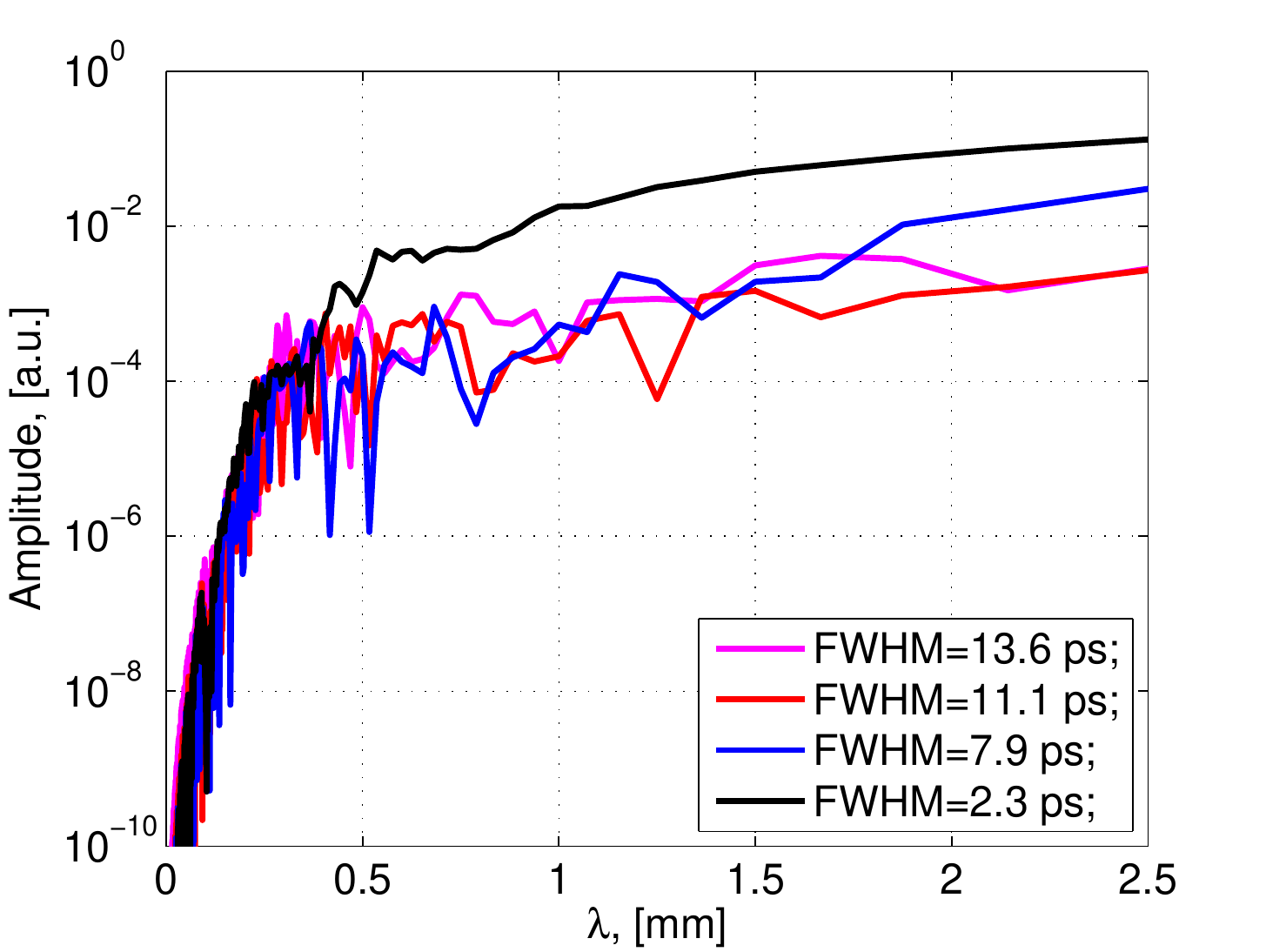}
  \caption{Form factor for the bunch profiles shown on figure~\ref{p123}. }
  \label{fig:FF}
\end{figure}

\begin{figure}[!htb]
  \centering
    \vspace*{-0.8cm}
    \includegraphics[width=0.8\linewidth]{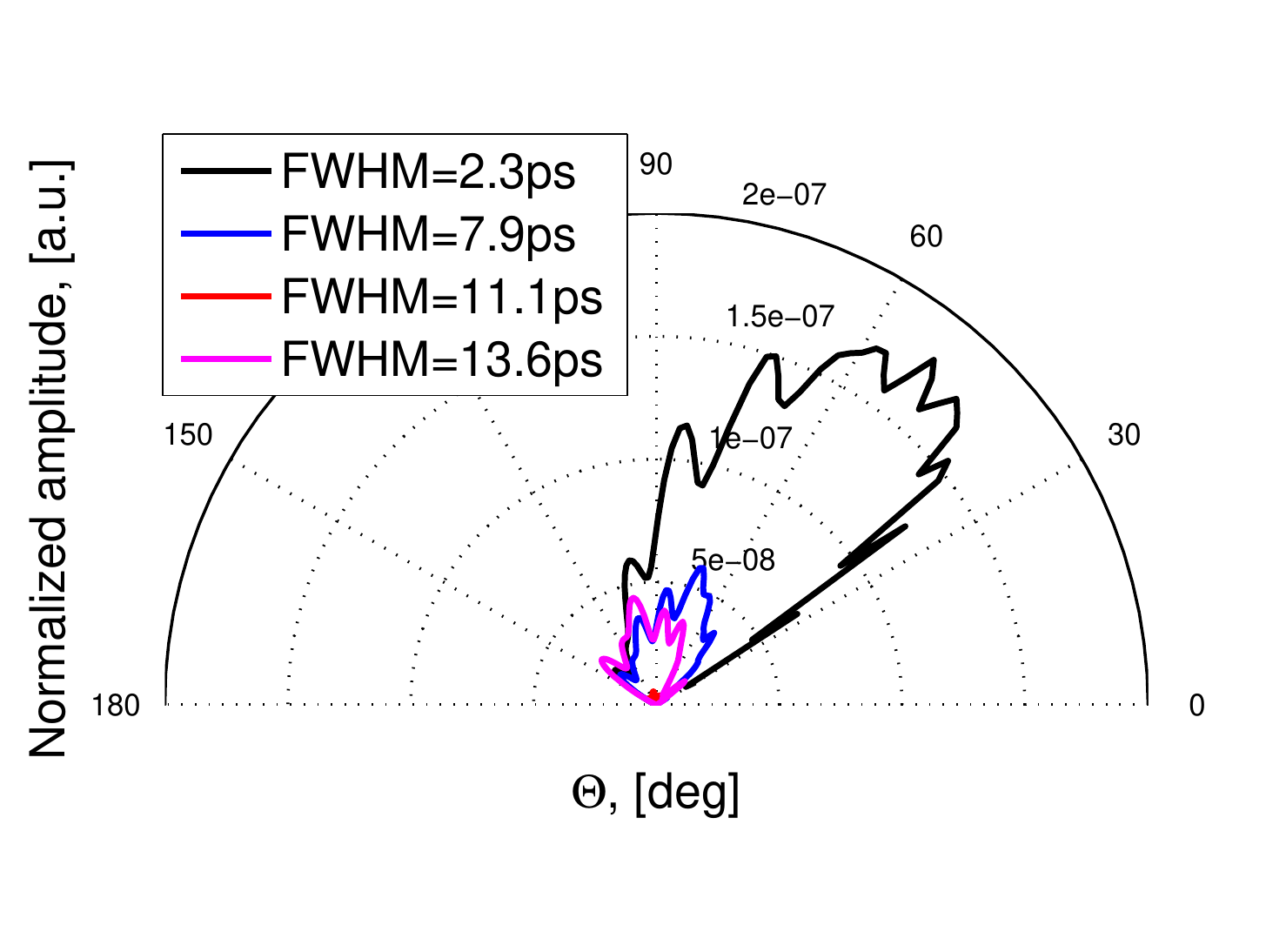}\\
    \vspace*{-0.8cm}
  \includegraphics[width=0.8\linewidth]{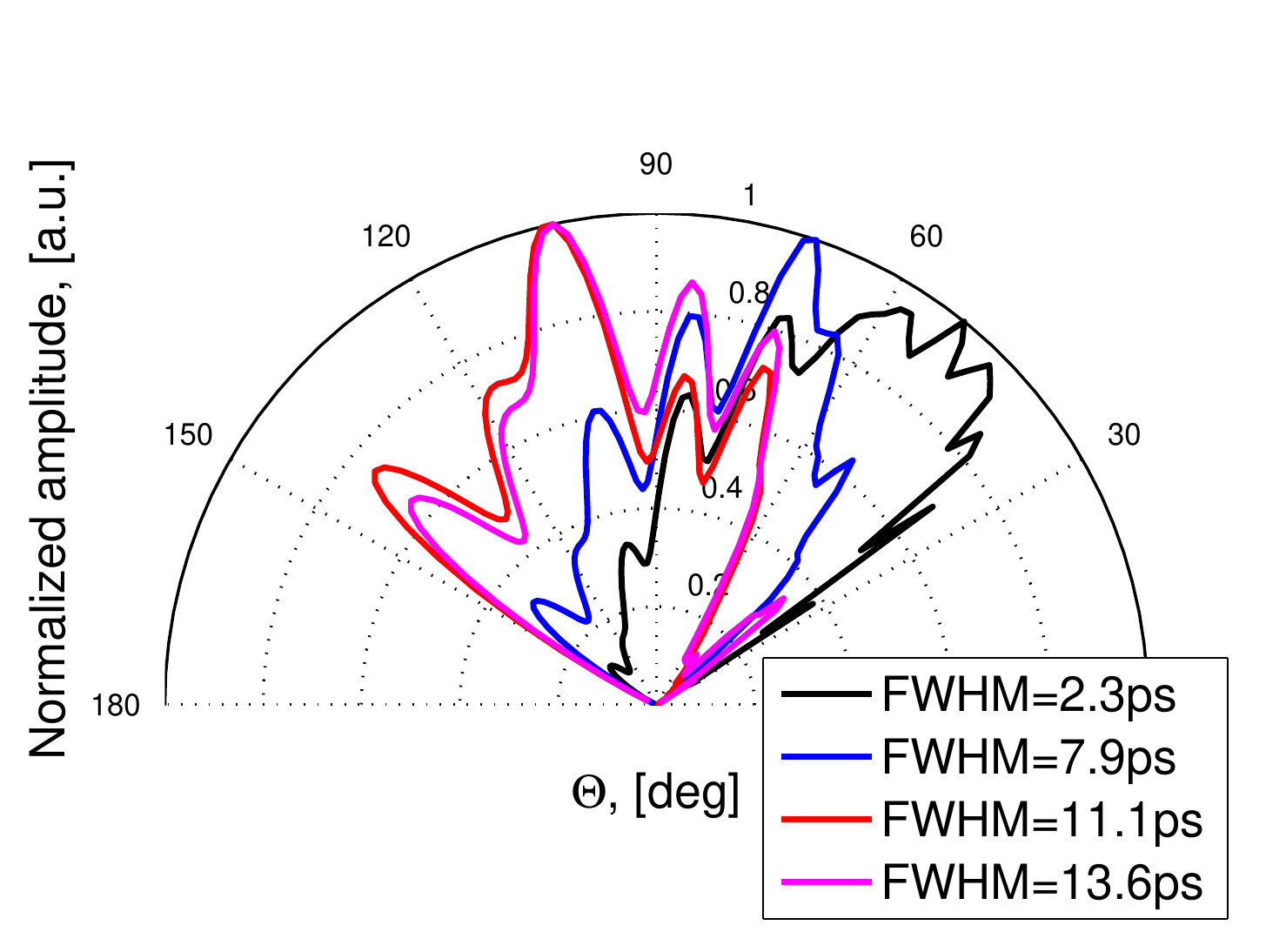}
  \caption{  Coherent Smith-Purcell spectrum as a function of the observation angle for different maximum field in the FB. The grating used for these simulations has a pitch of 8~mm and a blaze angle of $30^o$. The upper figure gives the energy distribution, the lower figure is normalized so that the maximum amplitude of each line is 1.}
  \label{sp}
  \label{sp-normalised}
  \end{figure}

On figure \ref{fig:FF} the form factor of the bunch at different AC phases is shown. On figure~\ref{sp} (top) one can see the predicted spectrum for CSPR (the code used is based on~\cite{gfw}). Making the distinction between these profiles to know if the bunch has the target length will be rather easy as it will be a matter of looking at the direction in which the signal is the most intense as shown on figure~\ref{sp-normalised} (bottom).

%\begin{figure}[!htb]
% \includegraphics[width=0.9\linewidth]{MOPMB005f6.pdf}

%  \caption{Coherent Smith-Purcell spectrum as function of observation angle for different maximum field in the FB. The grating used for these simulations has a pitch of 8 mm and a blaze angle of$30^o$.  }
%  \label{sp}
%\end{figure}

\begin{figure}[!h]
  \centering
  \vspace*{-0.4cm} 
   \includegraphics[width=0.8\linewidth]{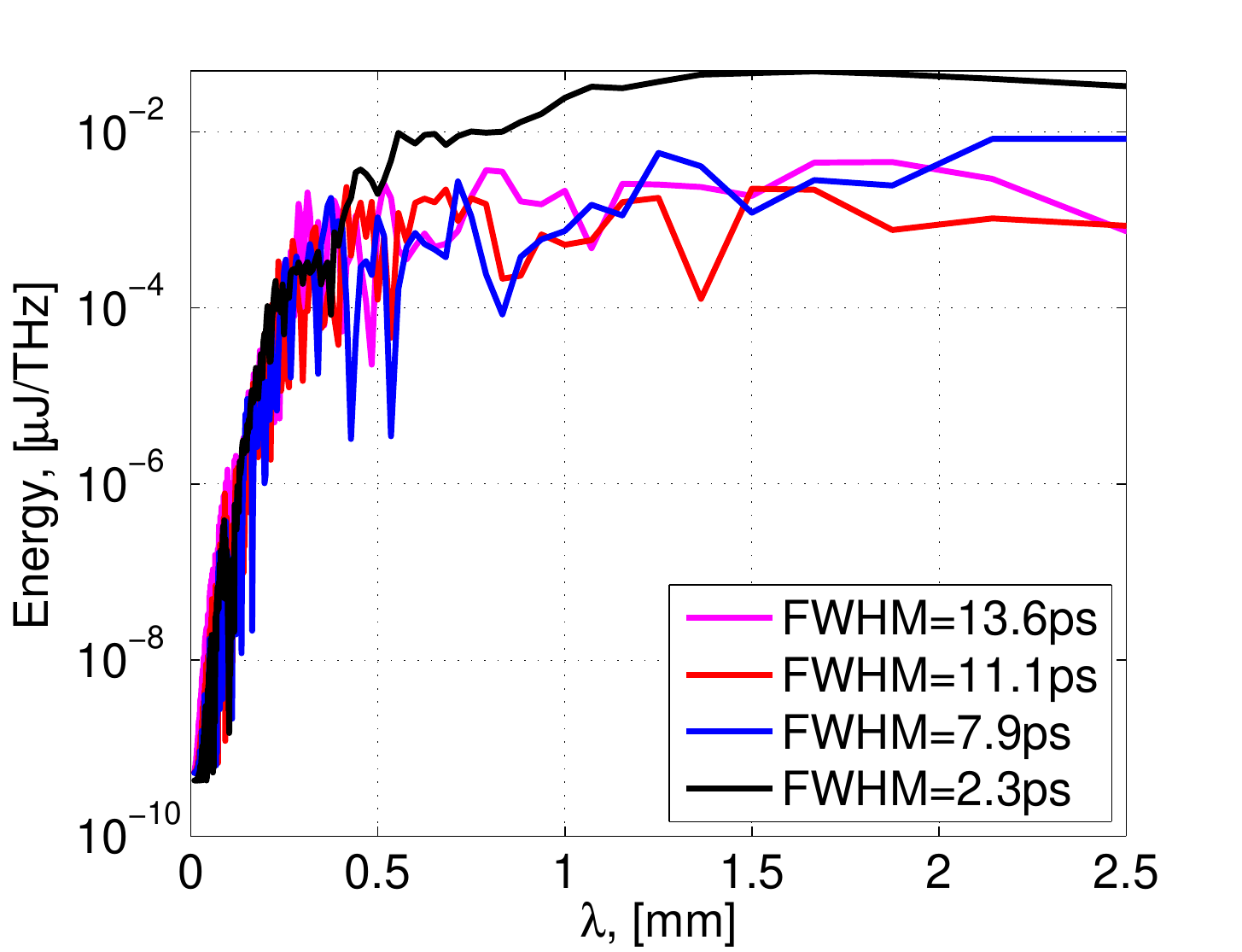} 
     \caption{CTR spectrums the different bunches profiles shown on figure~\ref{p123}.}
     \label{CTR-spectrum}
 \vspace*{-0.6cm}
\end{figure}

The CTR measurement will use a single detectors located at 90$^o$. The discrimination between pulse length will therefore be based primarily on CTR intensity but to enhance this phenomena we will also use band-pass filters in THz wavelength designed according to~\cite{filt}. Figure~\ref{CTR-spectrum} shows the CTR spectrum before filtering.%

Unlike the case of CSPR, CTR will only be used to estimate the bunch length and no detailed profile reconstruction will be attempted.%

%   \begin{figure}[!htb]
%  \centering
%  \includegraphics[width=0.9\linewidth]{plots/Filter2.eps}
 % \includegraphics[width=0.9\linewidth]{plots/CTR_Filter.eps}
%  \caption{Energy of CTR with  mesh filters.}
%  \label{filt}
% \end{figure}

\section{Conclusion}
We have simulated the CLIO Free Electron Laser using the ASTRA code. We have used these simulations to predict the longitudinal bunch length, its form factor and the expected signal yield using Coherent Smith-Purcell Radiation and Coherent Transition Radiation. The experimental chamber is currently being built and we expect our first experimental measurement in the coming few months.%

\null
%\begin{figure}[!htb]
% \centering
 % \includegraphics*[width=70mm]{THPME088f3.eps} \\
%  \includegraphics*[width=70mm]{THPME088f4.eps} 
%  \caption{$\Delta_{FWXM}$  (top) and $\chi^2$ (bottom) distribution of our 1000 simulations reconstructed using the Hilbert transform method.}
%   \label{profiles_stats_hilbert}
%\end{figure}

%\bibliographystyle{unsrt}
%\bibliography{biblio}

%%\begin{thebibliography}{9}   % Use for  1-9  references
%\begin{thebibliography}{99} % Use for 10-99 references

%\bibitem{accelconf-ref}
%	C. Petit-Jean-Genaz and J. Poole,
%	``JACoW, A service to the Accelerator Community,''
%	EPAC'04, Lucerne, July 2004, THZCH03,  p.~249,
%	\url{http://www.JACoW.org/e04/papers/THZCH03.PDF}

% \end{thebibliography}

\end{document}